# Field induced single molecule magnet behavior in Dy-based coordination polymer

Suraj Singh, Sheetal, Bandhana Devi, R.R. Koner, Aditi Haldar, and C. S. Yadav

School of Basic Sciences, Indian Institute of Technology Mandi, Mandi-175005 (H.P.), India

**Abstract:** - A new mononuclear Dysprosium based Coordination Polymer {Dy-CP} was investigated magnetically using dc and ac magnetic susceptibility. The dc magnetic susceptibility does not exhibit any long-range ordering down to 1.8 K and the negative value of Curie Constant (~ - 4 K) indicate the dominance of antiferromagnetic interactions between the Dy (III) spins. Ac susceptibility exhibits absence of single molecular magnet behavior at zero dc magnetic field and shows signal of quantum tunneling magnetization (QTM) below 8 K. However, on the superimposition of dc magnetic field (3 kOe), frequency dependent relaxation peak emerged at $T_f$ = 5 K and QTM signal suppress at higher fields. The intermediate value of Mydosh parameter calculated from the shift in peak position ($T_f$) in ac susceptibility reflects the formation of superparamagnetic state. Further, the temperature dependence of $T_f$ is analyzed with Arrhenius and Cole-Cole plot. The magnetic susceptibility analysis yields characteristics pre-relaxation factor $\tau_o$ =1.40 × 10$^{-12}$ sec and energy barrier $\Delta E/k_B$ = 93.4 K, indicating the slow spin relaxation. The Cole-Cole fit to the ac susceptibility data shows further evidence for the single ion spin relaxation. Thus, the magnetic measurements support the single-molecule magnet behavior in Dy-CP under the application of dc magnetic field.

**Introduction:**
The study of Single Molecule Magnets (SMMs) has received considerable attention due to their potential applications in quantum computing devices, large data storage and spintronics [1-5]. In particular, the lanthanide based coordination polymer molecules are found to be more fascinating candidates for the SMMs, due to their slow magnetic relaxation and large magnetic anisotropy owing to unquenched orbital angular momentum (L), high spin angular momentum (S) and large total angular momentum (J) values of the Lanthanide ions [6-10]. Both chemist and physicist has studied the luminescent nature of SMMs with proper correlation between magnetic and optical properties [11-13]. The interactions in the SMMs are attained by the overlap of bridging ligand orbital and the 4f orbitals of the rare-earth ions. Therefore the ligand designing is one of the essential components to achieve such interactions. Dy is one of the most used element for constructing SMMs [14,15] because of the possibility of magnetic data storage in single molecule at a temperature exceeding the liquid nitrogen temperature [10] .

The SMMs exhibits quantum tunneling magnetization (QTM) through the spin-reversal barrier via degenerate energy levels [16]. Though the spin relaxation mechanism is well known for transition metal based SMMs, it is still to be understood completely for polynuclear lanthanide based SMM systems. This is mainly because of the combined effect of several factors, such as high tunneling rates, weak magnetic interactions between the 4f ions and large magnetic anisotropy [17-20]. Therefore, understanding of the origin of spin relaxation observed in polynuclear 4f based SMMs remains an interesting challenge.

In this article we have presented the low temperature magnetic properties of the Dy based coordination polymer $C_{11}H_{18}DyN_3O_9$ (Dy-CP). The details of the compound synthesis and basic characteristic are already reported by Bandhana *et al.* [21]. Temperature and magnetic field dependent dc magnetization and ac susceptibility measurements have been performed using the Magnetic Property Measurement System (MPMS), from Quantum Design, USA in the temperature range 1.8 to 300 K. In Dy-CP, two Dy (III) ions interacts via two carboxylate oxygen atoms with distorted tricapped triangular prismatic geometry [21]. Our study shows evidences of strong anisotropy and the signal of quantum tunneling magnetization below 8 K. Ac susceptibility measurements in the presence of applied dc magnetic field confirm the slow magnetic relaxation behavior from Cole-Cole and Arrhenius fit with low value of $\alpha$ = 0.1 and energy barrier $\Delta E/k_B$ ~ 93 K. These features put Dy-CP on the edge of the Single-Molecule Magnets category for any reported Dy based mononuclear SMMs.

**Results and Discussions:**
The dc magnetic susceptibility (M/H) measurements of Dy-CP were performed under Zero Field Cooled (ZFC) and Field Cooled (FC) protocol in the presence of an applied field of 1 kOe and 3 kOe in the temperature range of 1.8-300 K. The temperature dependence of the dc magnetic susceptibility under 1 kOe (Figure 1) and 3 kOe field (Figure not shown) shows paramagnet like behavior with the complete overlapping of ZFC and FC curves, excluding the possibility of any type of long range magnetic ordering in the compound. We have fitted $1/\chi_{dc}$ versus T curve using Curie-Weiss equation (Figure 1(a)), $1/\chi = (T + \Theta_{CW})/C$; where $C = C_o g_J^2 J(J+1)/3$, $C_o = N_A\mu_o\mu_B^2/k_B = 4.71$ cm$^3$-K-mol$^{-1}$.

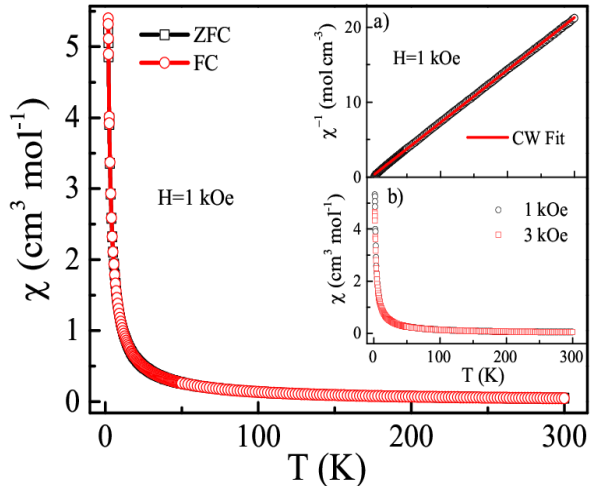
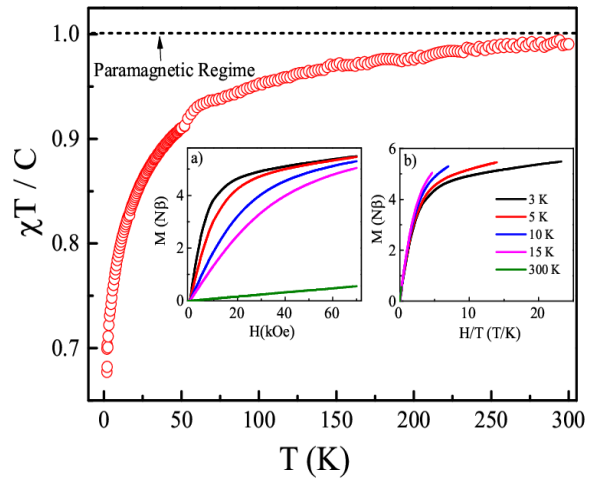

Figure 1: DC susceptibility curves of Dy-CP obtained under ZFC and FCC conditions at $H = 1$ kOe in the temperature range 1.8-300 K. Inset (a): $1/\chi$ vs T curve and red line the fit with Curie-Weiss Law. (b) $\chi$ vs T plot of Dy-CP at 1 kOe and 3 kOe under ZFC condition.

Figure 2: Temperature dependence of $\chi T/C$ for Dy-CP at T = 1 kOe. Inset (a) and (b) are Magnetic field response of magnetization and $M$ vs H/T plot for Dy-CP at different temperatures.

We obtained $\Theta_{CW} = -3.98$ K, $J \sim 15/2$, and $\mu_{eff} = 10.59$ $\mu_B$. The effective moment value is close to the theoretical value of 10.64 $\mu_B$ for Dy (III) ions. These results confirm that the Dy (III) ions solely contribute to the total magnetization. The negative value of $\Theta_{CW}$ points towards the antiferromagnetic exchange interaction between Dy (III) ions in the coordinated polymer compound. Figure 1(b) shows the dc magnetic susceptibility data (only ZFC) at 1 kOe and 3 kOe. The magnetic susceptibility overlap completely at both the fields indicates the absence of spin correlation with the application of magnetic field. The monotonic increase in the magnitude of dc susceptibility with decreasing temperature at 3 kOe without any observation of magnetic anomaly clearly indicates that the magnetic anomaly in ac susceptibility (discussed later) is not associated with any type of spin ordering.

Further to get the better understanding of the nature of interaction we have plotted $\chi T/C$ verse T in Figure 2. The magnitude of $\chi T/C$ is used to classify the nature of interaction between the magnetic ions [22]. If the systems exhibit ferromagnetic (FM) interactions then $\chi T/C > 1$ and in case of antiferromagnets (AFM) $\chi T/C < 1$. There exists a paramagnetic boundary between FM and AFM interaction regime for which $\chi T/C$ is equal to 1. It was observed that the system remains weakly antiferromagnetic throughout the investigated range with $\chi T/C < 1$. However, the presence of strong frustration prevents the possibility of long-range ordering in the system.

The isothermal magnetizations M versus H were performed at different temperatures (inset of Figure 2). At high temperature (T = 300 K), M(H) is nearly straight line, as expected for paramagnetic state. With the decrease in temperature, the curve develops a curvature, which is more pronounced at low temperatures. The M(H) curve doesn't saturate up to the field of 70 kOe and points to the presence of strong magnetic anisotropy [23]. At lowest measured temperature (T = 3 K), its maximum unsaturated magnetization value of 5.48 N$\beta$ for Dy-CP is almost half of the expected value (10 N$\beta$ for each Dy (III) ion with J = 15/2 and g = 4/3) [24]. Furthermore, to confirm the presence of magnetic anisotropy we plotted M versus H/T at different temperatures. As seen from the figure 2(b), there is discernable deviation at higher H/T value and all the curves do not superimpose on each other. This behavior indicates the presence of low-lying excited states and magnetic anisotropy in the compounds [23, 24].

In order to understand the magnetic relaxation behavior, we have measured the dynamic magnetization (ac susceptibility) of the compound in the low temperature range 1.8 - 50 K, and frequency range from 13 to 991 Hz, with the oscillation field of 3 Oe. At zero field, the real part of ac susceptibility ($\chi'$) exhibited frequency independent behavior (Figure 3a). However the out-of-phase (imaginary) part of the ac susceptibility ($\chi''$) shows a discernible change in behavior below 8 K (Figure 3(b)). The peak value is not obtained above 1.8 K but the observation of a tail of peak below 8 K is an indicative of fast QTM which is attributed to spin-reversal barrier via degenerate energy levels [25-27]. Such behavior is commonaly observed in lanthanides-based polymer systems [28-30]. The QTM in lanthanide systems is facilitated by the intermolecular interaction between lanthanide ions and poses a large energy barrier for magnetic tunneling [26, 31]. On the other hand, in Dy (III) ions the spin-parity effect reduces the appearance of QTM [25, 27, 32]. The intensity of the tail increases at higher frequencies which clearly demonstrate that QTM in Dy-CP is consistent with the Lanthanides SMMs shows the single-ion relaxation behavior [31].

Later, we have done the ac susceptibility measurements in the presence of dc field. The ac susceptibility of Dy-CP measured under 3 kOe dc magnetic field is shown in figure 3(c) As seen from the figure, we observed frequency dependence at low temperature with a clear peak near 5 K. The shift in the peak position with varying frequency indicate the presence of glassy state in the system [33]. For this we have calculated Mydosh parameter $\varphi$ by using the relation [33]

$$\varphi = \frac{\Delta T_f}{T_f \Delta(\log f)} \quad \ldots\ldots (1)$$

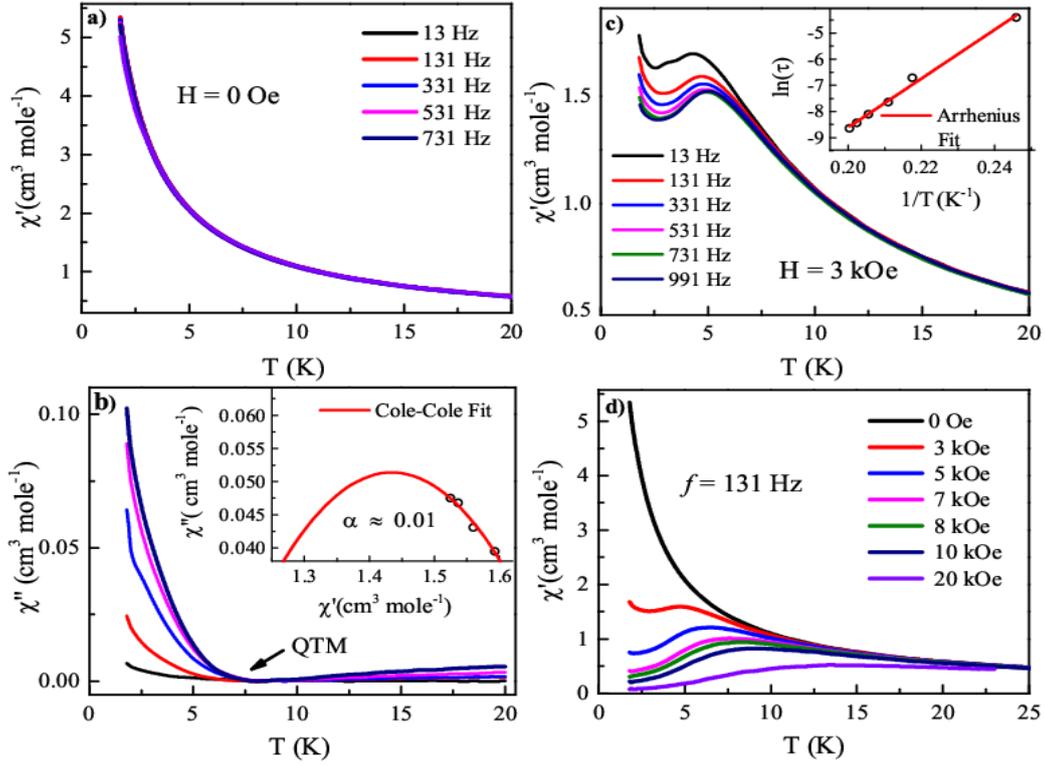

Figure 3: (a) & (b) Temperature dependence of $\chi'$ and $\chi''$ at H = 0 Oe performed at various frequencies. (c) $\chi'$ vs T at H = 3 kOe between 13 Hz to 991 Hz. (d) Temperature dependence of $\chi'$ at $f$ = 131 Hz perofrmed at various dc applied fileds. Inset of (b) and (c) reperensets the Cole-Cole plot and Arrhenius fit using equation 2 and 3.

Where $T_f$ is the transition temperature at lowest measured frequency and $\Delta T_f$ is the difference between the transition temperature measured at the lowest and highest frequency. By using the peak temperature, maximum and minimum frequency values, we obtained $\varphi = 0.1$, as predicted for superparamagnet, which reject the presence of the spin glass behaviour. Further, the frequency dependence of ac susceptibility data was analysed by the Cole-Cole plot using equation [34, 35]

$$\chi(\omega) = \chi_S + \frac{\chi_T - \chi_S}{1 + (i\omega\tau)^{1-\alpha}} \quad \ldots\ldots (2)$$

Where $\chi_T$ is the isothermal susceptibility and $\chi_S$ is the adiabatic susceptibility and $\alpha$ ranges between $0 \leq \alpha \leq 1$. $\alpha = 0$ indicates no spread in the relaxation time and deviation from this value indicates wide distribution of relaxation time due to the formation of clusters [33-38]. To obtain $\alpha$, we did the fitting using generalized Debye model equation (2) in the Cole-Cole plots of $\chi''$ vs $\chi'$ for Dy-CP under 3 kOe applied dc field. We found the value of $\alpha \sim 0.01$ at 5 K. A very low value of $\alpha$ for Dy-CP corresponds to the narrow distribution of the relaxation time. For the relaxation time of Dy-CP, the frequency dependent transition temperature in the range of 13 – 991 Hz is fitted with the Arrhenius equation [34]

$$\tau = \tau_0 \exp(\Delta E / k_B T) \quad \ldots\ldots (3)$$

The obatained energy barrier ($\Delta E/k_B$) of 93.4 K and pre-relaxation factor $\tau_o$ of $1.40 \times 10^{-12}$ sec (inset of Figure 3(c)), correspond to the single molecule magnets range ($10^{-6} – 10^{-12}$ sec) [39, 40]. This energy barrier is larger than any previously reported data for mononuclear Dy ions single molecule magnets. These observations suggest that the Dy-CP MOF fall in the category of the single molecule magnets.

Figure 3(d) shown for $f$ = 131 Hz frequency, depicts the dependence of dynamic magnetization on applied dc magnetic field more clearly. The ac susceptibility at zero field shows monotonous Curie like behavior, and the application of dc field assists in arresting of the spin dynamics at low temperatures by giving rise to peak shape anomaly, which gets broadens for higher fields. The interesting observation is the appearance of frequency dependent peak at an applied dc field of 3 kOe, indicating SMM behavior (Figure 3c). A relaxation peak was observed below 8 K with a tail of

QTM, which is now shifted to low temperatures on applying dc field. On further increase in dc applied field H > 5 kOe, only a single-relaxation peak is observed without the presence of a tail, confirming the complete suppression of QTM. Essentially, we found that QTM can be blocked with a suitable applied dc magnetic field. The application of field lifts the degeneracy and prevents the tunneling of electrons from +$M_S$ state to -$M_S$ and results in the suppression of QTM.

**Conclusion:** Dy-CP was characterized magnetically and indicates the presence of weak antiferromagnetic interaction. Reduced value of saturation magnetization indicates the presence of low-lying excitation and large anisotropy in the system. Ac susceptibility shows paramagnetic like feature different from other members of SMMs group. Interestingly, imaginary part of ac susceptibility shows signature of QTM below 8 K which further suppressed with the application of dc magnetic field due to the splitting of degenerate energy levels. Apart from this, field induces SMMs behavior in Dy-CP at T = 5 K, confirmed from the Arrhenius and Cole-Cole fit. Our work presents Dy-CP as an interesting candidate in SMMs class and can be further studied magnetically by advanced measurements.

**Acknowledgement:** We thank AMRC IIT Mandi for the experimental facility and IIT Mandi for the financial support.